\documentstyle[prl,aps,multicol,epsf]{revtex}
\input psfig
\begin{document}
\title{ Pumping at resonant transmission and transferred charge quantization}
\author{Y. Levinson }
\address{ Department of Condensed Matter Physics, The Weizmann
Institute of Science, Rehovot 76100, Israel}
\author{O. Entin-Wohlman}
\address {School of Physics and Astronomy,
Raymond and Beverly Sackler Faculty of Exact Sciences,\\
 Tel Aviv University, Tel Aviv 69978, Israel}
\author { P. W\"olfle}
\address{Institut f\"ur Theorie der Kondensierten Materie,
Universit\"at Karlsruhe, 76128 Karlsruhe, Germany}
\date{\today}
\maketitle
\begin {abstract}
We consider pumping through a small quantum dot  separated from the leads
by two  point contacts, whose conductances, $G_{1}$ and $G_{2}$, serve as
pumping parameters. When the dot is pinched, i.e.
$G_{1},G_{2}\ll e^2/h$, we find that there is a ``resonance line"
in the parameter plane $\{G_{1}, G_{2}\}$
along which the Fermi energy in the leads aligns with the energy of the
quasi-bound state in the quantum dot. When
$G_{1}$ and $G_{2}$ are modulated periodically and adiabatically such that
the pumping contour defined by $G_{1}=G_{1}(t)$ and $ G_{2}=G_{2}(t)$
encircles the resonance line, the current is quantized:
the charge pumped through the dot during each period of the modulation is close
to a single electronic charge.
\end {abstract}
\pacs {PACS numbers: } \vspace{-6mm}
\begin{multicols}{2}
The question of current quantization was first addressed in Ref. \cite{Th83}.
It was shown that under certain conditions the current $J$ induced by a
slowly moving periodic potential profile corresponds to an integral
number of
electronic charges $e$ transferred through the sample cross-section during
a single {\it temporal} period $T$ of the perturbation:
$J=(e/T)\times$integer, where $T=L/v$, with $L$ and $v$ being the period
of the potential profile and its velocity.
 Current quantization was
observed in a quasi-one-dimensional (1D) GaAs channel \cite{Sh96,Ta97},
in which the potential profile had been created
by the piezoelectric potential of a surface
acoustic wave. A 1D model describing charge quantization in this
experiment was discussed in Ref. \cite{Ma00}. It was demonstrated
\cite{LEW}
that the current induced by a moving potential profile is equivalent to
pumping, a phenomenon \cite{AG99} which had been first considered in Ref.
\cite{He91} and has attracted recently much theoretical
\cite{Br98,AA98,SAA00,S00} and experimental \cite{Sw99,Po92,Kou91} interest.

 The pumping current is excited not by applying a voltage difference, but
by periodically changing some properties of the system (i.e., parameters of
the system Hamiltonian), for example, the confining potential in a
nanostructure. The fingerprint of pumping appears when the  frequency
$\omega$ is smaller than any characteristic energy of the system
(adiabatic pumping). Then the charge transferred during a single period,
$Q=J\times (2\pi/\omega)$, is independent of $\omega$. However,
this charge is {\it not necessarily quantized}.

The simplest case is pumping at zero temperature through a two-terminal
device with single-channel terminals. When two parameters of the device are
modulated periodically the charge transferred by spinless electrons is
\cite{Br98}
\begin{equation}
\label{Q}
 Q={e\over\pi}\int\int d\lambda_{1}d\lambda_{2}\;\;\Pi(\lambda_{1},\lambda_{2}),
\end{equation}
with
\begin{equation}
\label{Pi}
 \Pi=-{\rm Im}\left[{\partial s_{11}\over\partial \lambda_{1}}
{\partial s_{11}^{*}\over\partial \lambda_{2}}+{\partial
s_{12}\over\partial \lambda_{1}}{\partial s_{12}^{*}\over\partial
\lambda_{2}}\right].
\end{equation}
Here $s_{\alpha\beta}$ is the scattering matrix of the device calculated
at the Fermi energy $\epsilon_{F}$ in the leads and
$\lambda_{1}=\lambda_{1}(t),\lambda_{2}=\lambda_{2}(t)$ are adiabatically
modulated parameters, displaying a closed ``pumping contour"
(counterclockwise)
 in the parameter plane $\{\lambda_{1},\lambda_{2}\}$. The integration
is over the area  encircled by the pumping contour. It is obvious from
this result that $Q$ is quantized only under special circumstances.

The quantization of the charge pumped through a {\it large, almost open}
quantum dot (QD) with vanishing level spacing was analysed in Refs.
\cite{AA98,SAA00}. In this letter we consider the opposite case, of a {\it
small, strongly pinched} QD, supporting resonant transmission. We show
that charge quantization can be achieved at zero temperature, provided
that the pumping contour is properly chosen.

Our results can be formulated in the following generic way. Consider a QD
in a two-dimensional electron gas (2DEG), separated from the leads by two
single-channel point contacts PC1 and PC2 with conductances $ G_{1}$ and
$G_{2}$ controlled by split-gate voltages $U_{1}$ and $U_{2}$. These
conductances will serve as pumping parameters. (A device of this type was
used, for example, in the experiments of Ref. \cite{Ko91}). Let the Fermi
level in the 2DEG $\epsilon_{F}$ be close to $\epsilon_{0}$, a level which
is formed in the isolated QD, when $G_{1},G_{2}=0$. When the QD is not
isolated, but is strongly pinched, $G_{1},G_{2}\ll 1$ (for conductances in
units of $e^2/h$ ), the level $\epsilon_{0}$ turns into a resonance at
$\epsilon_{0}+\Delta\epsilon_{0}+i\Gamma$, with $\Delta\epsilon_{0}<0$ and
$\Gamma\ll |\Delta\epsilon_{0}|$. Then, when $\epsilon_{F}<\epsilon_{0}$,
there is a ``resonance line" in the parameter plane $\{G_{1},G_{2}\}$,
where the resonance energy is aligned with the Fermi level,
$\epsilon_{F}=\epsilon_{0}+\Delta\epsilon_{0}$. Along this line the
transmission through the QD is at resonance. The conductance
$G(G_{1},G_{2})$ of the QD is then sharply peaked (the width of the peak
is determined by $\Gamma$) in the direction perpendicular to the resonance
line, and decreases towards the ends of this line. The transmission is
maximal at $G_{1}=G_{2}$ for a symmetric QD. We find that the pumped
charge is quantized, $Q\approx e$, and $|Q-e|\rightarrow 0$ when
$\epsilon_{F}$ approaches $\epsilon_{0}$ from below, provided that the
pumping contour encircles the resonance line.

The resonance line can be found experimentally. When PC1 is pinched and
PC2 is open, the conductance of the QD, $G$, is dominated by the conductance
of PC1, i.e. $ G=G_{1}$. Measuring the dependence of $G$
on $U_{1}$ yields
the relation between $G_{1}$ and $U_{1}$. In a similar way one finds the
relation between $G_{2}$ and $U_{2}$. Using these relations and measuring
$G$ as  function of $U_{1}$ and $U_{2}$,  the resonance line can be obtained.

To derive the above  result we consider a 1D model, similar to the one
used in Ref. \cite{He91} (and also in Ref. \cite{W00}), in which the QD is
confined by two potential barriers, located at $x=\mp a/2$. The scattering
matrix of the QD,
 $ \left|\begin{array}{cc}s_{11}&s_{12}\\s_{21}&
 s_{22}\end{array}\right|,$
(1 and 2 denote the terminals at $x=-\infty $
and $x=+\infty$, repsectively)
can be constructed from the scattering matrices of the
two barriers:
\begin{eqnarray}
\label{s}
s_{12}=s_{21}&=&t_{1}t_{2}\sigma/D,\nonumber\\
s_{11}=r_{1}+t_{1}^{2}r_{2}\sigma^{2}/D,\ \ &&\ \
s_{22}=r_{2}+t_{2}^{2}r_{1}\sigma^{2}/D,
\end{eqnarray}
where $r_{1,2}$ and $t_{1,2}$ are the reflection and the transmission
amplitudes of the two  barriers,
 $\sigma=e^{ika},\; D=1-r_{1}r_{2}\sigma^{2}$, and
$k=(2m\epsilon)^{1/2}$. The phase reference points for the waves on the
left and on the right of the QD are at $x= \mp a/2$, respectively.

Let us confine ourselves for simplicity to delta-function potential
barriers,
 $V_{1,2}\delta (x\mp a/2)$. In this case,
\begin{equation}
\label{trv}
 t_{1,2}=( 1+imV_{1,2}/k)^{-1},\qquad
r_{1,2}=t_{1,2}-1,
\end{equation}
and the scattering matrix elements are
\begin{eqnarray}
\label{sxi}
 s_{11}&=&[1-i\xi_{2}-(1+i\xi_{1})\sigma^2]/D',\nonumber\\
s_{22}&=&[1-i\xi_{1}-(1+i\xi_{2})\sigma^2]/D',\nonumber\\
s_{12}&=&\xi_{1}\xi_{2}\sigma/D',\ \
D'=-(1-i\xi_{1})(1-i\xi_{2})+\sigma^2,
\end{eqnarray}
with $\xi_{1,2}=k/mV_{1,2}$.

The energies $\epsilon_{0}$ of the bound states in the isolated QD, (that
is, when $\xi_{1}=\xi_{2}=0 $), are  given by
 $\exp (2ik_{0}a)=1$ with  $\epsilon_{0}= k_{0}^2/2m.$
Below we  assume the Fermi energy  to lie in the vicinity of one of these
levels,  $ \delta\equiv (\epsilon_{F} -\epsilon_{0})/ (v_{0}/2a)\ll 1$,
where $v_{0}= k_{0}/m$. Here $\delta$ is the detuning of the Fermi energy
from the bound state, measured in units of the level spacing in the
isolated QD.

At near resonance Fermi energy, the smooth energy dependence of $t_{1,2}$
and $r_{1,2}$ can be ignored, putting $k=k_{0}$, i.e., $\xi_{1,2}=
k_{0}/mV_{1,2}$. The latter will serve as the pumping parameters. Note
that for a pinched QD these parameters are  related to the conductances of
the point contacts in a simple way: $G_{1,2}=\xi_{1,2}^2$. We  assume that
during the pumping cycle the barriers remain high. That is, the QD will
support resonant transmission during the whole pumping cycle.

For finite, but high, barriers  ($\xi_{1},\xi_{2}\ll 1$), the bound
state of energy $\epsilon_{0}$ turns into a quasi-bound state
with complex energy
$\epsilon=\epsilon_{0}+\Delta\epsilon_{0}+i\Gamma$, obtained when
$D'$ in Eq. (\ref{sxi}) tends to zero. One finds
\begin{eqnarray}
\label{re}
 \Delta\epsilon_{0}= -(v_{0}/
2a)(\xi_{1}+\xi_{2}),\qquad \Gamma=(v_{0}/2a)\xi_{1}\xi_{2}.
\end{eqnarray}
The quasi-bound state lies always below the corresponding bound one and
its width $\Gamma$ is much smaller that the shift $\Delta\epsilon_{0}.$
Resonant transmission through the QD  occurs when $\epsilon_{F}$ is
aligned with the quasi-bound state energy $\epsilon_{0}+\Delta\epsilon_{0}$
up to the level width $\Gamma$. Since $\Gamma\ll |\Delta\epsilon_{0}|$
this is possible only when $\epsilon_{F}$ is below $\epsilon_{0}$, i.e.
$\delta<0$. Hence, only this case will be considered.

The alignment of the quasi-bound state energy with the Fermi level determines
the resonance line
in the parameter plane:  $\xi_{1}+\xi_{2}=|\delta|$.
Examination of Eq. (\ref{sxi}) reveals that when $|\delta|\ll 1$ the
conductance of the QD $G=|s_{12}|^2$ is a sharp function along the direction
perpendicular to this resonance line, with width $\delta^2$, corresponding
to the narrow resonance in the QD. Along the resonance line $G$ has its
maximum for a symmetric QD, at $\xi_{1}=\xi_{2}$, decreasing towards its
ends. \setlength{\unitlength}{7cm}
 \vspace{-17mm}
\begin{figure}
\begin{picture}(1,1)
\put(0,0){\psfig{figure=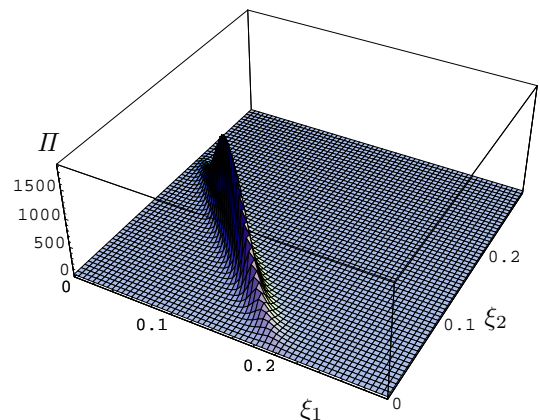,width=1\unitlength}}
\put(.55,-.005){\makebox(0,1)[lb]{$\xi_{1}$}}
\put(.9,.17){\makebox(0,1)[lb]{$\xi_{2}$}}
\put(.05,.50){\makebox(0,1)[lb]{${\it\Pi}$}}
\end{picture}
\caption{$\it\Pi$ as a function of $\xi_{1}$ and $\xi_{2}$ for
$\delta=-0.2$} \label{Pi1}
\end{figure}

For the pumping parameters $\xi_{1},\xi_{2}$ the function $\Pi $ defined in
Eqs. (\ref{Q}) and (\ref{Pi}) is given by
\begin{eqnarray}
\Pi(\xi_{1},\xi_{2})&=&M(\xi_{1},\xi_{2})/N^2(\xi_{1},\xi_{2}),\nonumber\\
M(\xi_{1},\xi_{2})&=&\xi_{1}\xi_{2}\;[\xi_{1}\xi _{2}\sin\delta+(
\xi_{1}+\xi_{2})(1-\cos\delta)],\nonumber\\
N(\xi_{1},\xi_{2})&=&
|D'|^2=\xi_{1}^2\xi_{2}^2+(\xi_{1}+\xi_{2})^2+2(\xi_{1}+\xi_{2})\sin\delta
\nonumber\\
&+&  2(1-\xi_{1}\xi_{2})(1-\cos\delta).
\end{eqnarray}
Similar to the conductance
$ G=|s_{12}|^2=\xi_{1}^2\xi_{2}^2/N$, the function $\Pi$
 has its  maximum near the resonance line, see
Fig. \ref{Pi1}. Because of this sharp maximum, the resulting integral is
not sensitive to the form of the integration contour, when the  contour
[cf Eq. (\ref{Q})] in the parameter plane $\{\xi_{1},\xi_{2}\}$ encircles
the resonance line  at distances larger than $\delta^2$. The main
contribution to the integral comes from that part of the area where
$\xi_{1},\xi_{2}\ll 1$, i.e. where the QD is strongly pinched and the
transmission through it is resonant.

To calculate the integral and to obtain the pumped charge,
it is convenient to substitute
$\xi_{1}+\xi_{2}=|\delta| p, \; \xi_{1}-\xi_{2}=|\delta| pz,\;
(0<p<\infty,\; -1<z<+1).$ For $|\delta|\ll 1$  we then have
\begin{eqnarray}
M=\delta^5 f(p,z),\qquad N=\delta^2 [(p-1)^2+\delta^2 g(p,z)],\nonumber\\
 f(p,z)=-{1\over 16}p^3(1-z^2)[2-p(1-z^2)],\nonumber\\
 g(p,z)=-{1\over
12}+{1\over 3}p-{1\over 4}p^2(1-z^2)+{1\over 16}p^4(1-z^2)^2.
\end{eqnarray}
The pumped charge is now given by
\begin{equation}
Q={e\over 2\pi}\delta^3\int pdp\int dz {f(p,z)\over
[(p-1)^2+\delta^2g(p,z)]^2}.
\end{equation}
The integration over $p$ is  performed using
the formal relation
$(x^2+\eta^2)^{-2}=(\pi/ 2\eta^3)\delta (x)$  for $\delta\rightarrow 0$,
with the result
\begin{equation}
\label{qint}
 Q=e\int dz {1-z^4\over (1+z^2)^3}.
\end{equation}
When the contour encircles the whole resonance line the integration over
$z$ is from $-1$ to $+1$, resulting in $Q=e$, i.e. the pumped charge is
quantized. In fact, the contour cannot encircle the whole resonance line
{\it exactly} since its ends $z=\pm 1$ correspond to infinite barriers.
However, these ends contribute little to the integral (\ref{qint}).
  \narrowtext
\setlength{\unitlength}{4cm}
 \begin{figure}
 \begin{center}
\begin{picture}(1,1)
\put(0,0){\psfig{figure=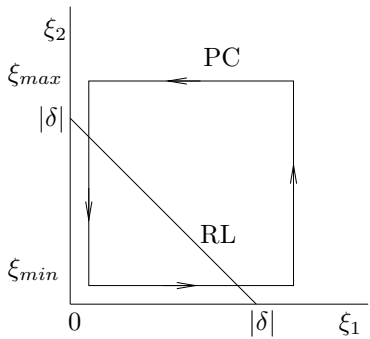,width=1\unitlength}}
\put(0,-.08){\makebox(0,1)[lb]{$0$}}
\put(.9,-.08){\makebox(0,1)[lb]{$\xi_{1}$}}
\put(-.20,.1){\makebox(0,1)[lb]{$\xi_{min}$}}
\put(-.20,.75){\makebox(0,1)[lb]{$\xi_{max}$}}
\put(.6,-.08){\makebox(0,1)[lb]{$|\delta|$}}
\put(-.08,.9){\makebox(0,1)[lb]{$\xi_{2}$}}
\put(-.1,.6){\makebox(0,1)[lb]{$|\delta|$}}
\put(.44,.21){\makebox(0,1)[lb]{RL}}
 \put(.45,.8){\makebox(0,1)[lb]{PC}}
 \end{picture}
 \end{center}
 \caption{Resonance line and pumping contour}
 \label{pc}
 \end{figure}
To examine the quantization accuracy,
we have computed
$q\equiv (1-Q/e)\times 100$\% in the case where the pumping contour in the
$\{\xi_{1},\xi_{2}\}$ plane is a square box, as shown in Fig. \ref{pc}.
Fig. \ref{q} exhibits $q$ as  function of $\xi_{min}$ and $\xi_{max}$
for $\delta=-0.2$.

The pumping contour in Fig. \ref{pc} corresponds to the conductances $G_{1}$
and $G_{2}$ being  alternatively modulated between $G_{min}=\xi_{min}^2$
and $G_{max}=\xi_{max}^2$, such that  the periodic functions
$G_{1}(t)$ and $G_{2}(t)$ are shifted by a quarter of a period, which
corresponds to the maximal possible pumping by a small perturbation \cite{Br98}.

>From Fig. \ref{q} it is seen that the accuracy of the quantization is not
very sensitive to $\xi_{max}$ provided that $\xi_{max}>|\delta|$. It is
more sensitive to the value of $\xi_{min}$. In terms of the conductances,
this means the following. The maximal resonance conductance of the pinched
QD,  for a given $\epsilon_{F}$, occurs at the center of the resonance
line where  $G_{1}=G_{2}\equiv G_{0}={1\over 4}|\delta|^2$. To achieve
high levels of accuracy of
 quantization one has to choose
$1\gg G_{max}>4G_{0}$ and make $G_{min}$  as small as possible. For
example, for $|\delta|=0.2$, which is equivalent to $G_{0}=10^{-2}$, one
obtains $q=0.8\%$, when $G_{max}=6\times 10^{-2}$ and $G_{min}=10^{-4}$.
Note that in the experiments \cite{Ko91} the maximal resistance of the PC is
about 100 G$\Omega$, which corresponds to $G_{min}\simeq 10^{-7}$, while
the minimal resistance is about 1 M$\Omega$, which corresponds to
$G_{max}\simeq 10^{-2}$. It follows that accuracy  much
higher than $1\%$ is accessible in the experiments.
 \vspace{-25mm}
 \setlength{\unitlength}{8cm}
\begin{figure}
 \begin{center}
\begin{picture}(1,1)
\put(0,0){\psfig{figure=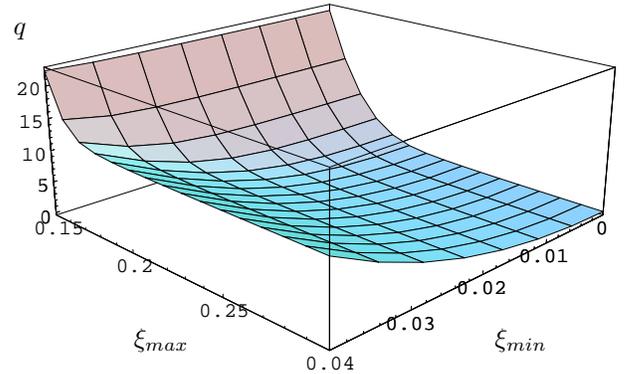,width=1\unitlength}}
\put(.2,.05){\makebox(0,1)[lb]{$\xi_{max}$}}
\put(.8,.05){\makebox(0,1)[lb]{$\xi_{min}$}}
\put(0,.57){\makebox(0,1)[lb]{$q$}}
\end{picture}
 \end{center}
 \caption{Accuracy of charge  quantization (in percents) as a function
  of $\xi_{max}$ and $\xi_{min}$ for $\delta=-0.2$}
\label{q}
\end{figure}

Even though our model is simplified, it serves to demonstrate that the
basic concepts of the quantization, namely the resonant transmission due
to a quasi-bound level in the QD  and the existence of a resonance line in
the parameter plane, are not very sensitive to details of the model chosen
for the barrier potentials. This strongly supports our generic formulation
of the conditions for charge quantization.

It has been shown in Ref. \cite{SAA00} that charge quantization is related
to the topological properties of phases of the scattering matrix. The
argument is based on the representation of the scattering matrix in terms
of the conductance $G$ and two phases $\alpha$ and $\beta$,
\begin{equation}
 \label{smQDp}
s= e^{i\phi} \left|\begin{array}{cc}(1-G)^{1/2}e^{i\alpha}&iG^{1/2}
  \\iG^{1/2}&(1-G)^{1/2}e^{-i\alpha} \end{array}\right|.
\end{equation}
(We have allowed here for an overall phase factor, $e^{i\phi }$, which is
missing in Ref. \cite{SAA00}.) It can be checked, using the unitarity of
the scattering matrix, that $\Pi$ in Eq. (\ref{Pi}) is invariant with
respect to $s\rightarrow e^{i\phi}s$. Using this result and
Stoke's theorem the pumped charge can be represented as
an integral along the pumping contour \cite{SAA00}
\begin{eqnarray}
Q={e\over 2\pi}\oint ds (1-G){\partial\alpha\over\partial s}.
 \label{ci}
\end{eqnarray}
The phase $2\alpha=\arg s_{11}- \arg s_{22}$ is shown (in units of $\pi$)
in Fig. \ref{pd}. This phase jumps by $\pm 2\pi$ at the resonance line, on
both sides of its center, respectively,  and hence {\it cannot} be
expressed as a continuous, single-valued function of $\xi_{1}$ and
$\xi_{2}$. The reason for this is that at the center of the resonance line
the reflection amplitudes $s_{11}=s_{22}=0$ and their phases $\arg s_{11}$
and $\arg s_{22}$ are not defined. By contrast, $\arg s_{12}$ {\it can} be
presented as a continuous, single-valued function since the transmission
amplitude $s_{12}\neq 0$ everywhere.

The contribution to the integral Eq. (\ref{ci}) comes from the jumps of
$\alpha$ at the resonance line. Alternatively, when the phase $\alpha$ is
required to be continuous on a selected contour which encircles the
resonance line, it will change by $2\pi$ upon closing the contour. It
follows that the condition for charge quantization formulated in Ref.
\cite{SAA00},that  {\it  the conductance $G$ has  to be small on the
pumping contour}, is not sufficient. Our results imply that for the charge
to be quantized {\it and not to be zero} the pumping contour has to
encircle the resonance line, where {\it the conductance $G$  is large}.
 \vspace{-2cm}
 \setlength{\unitlength}{8cm}
\begin{figure}
 \begin{center}
\begin{picture}(1,1)
\put(0,0){\psfig{figure=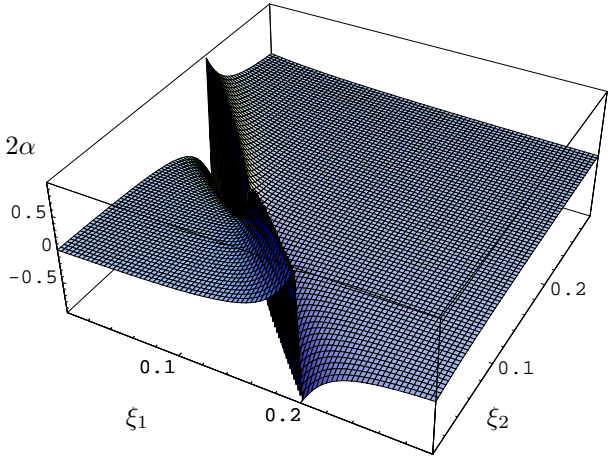,width=1\unitlength}}
\put(.2,.05){\makebox(0,1)[lb]{$\xi_{1}$}}
\put(.8,.05){\makebox(0,1)[lb]{$\xi_{2}$}}
\put(0,.5){\makebox(0,1)[lb]{$2\alpha$}}
\end{picture}
 \end{center}
 \caption{Reflection phase difference}
\label{pd}
\end{figure}

It is worthwhile to note that although the center of the resonance line
may appear as a branching point, this is not so: from Eq. (\ref{sxi}) it
can be seen that the matrix elements $s_{ik}$ are not analytic functions
of $\xi_{1}+i\xi_{2}$.

The quantization can be easily modified by finite temperature effects.
Then the charge, $Q$, has to be averaged over $\epsilon_{F}$ within the
temperature interval. A finite temperature will affect the charge
quantization when $T\gtrsim |\epsilon_{F}-\epsilon_{0}|$. To estimate this
effect consider the QD used in the experiments of Ref. \cite{Ko91}. The
level spacing in the QD  is 0.03 meV, which for $|\delta|=0.2$ corresponds
to detuning $|\epsilon_{F}-\epsilon_{0}|=6\mu$eV = 70 mK. Temperatures of
this order will destroy the quantization.

The transferred charge quantization discussed here is different from the
quantization observed in turnstile-type devices \cite{Po92,Kou91}. First,
the quantization in such devices is essentially based on the Coulomb
blockade, fixing the number of electrons in the  QD's (or quantum boxes).
Second, one can see from Eqs. (\ref{Q}) and (\ref{Pi}), that  the {\it
phases} of the scattering matrix are crucial for the appearance of the
transferred charge $Q$, while the turnstile device operation is usually
described in terms of sequential transitions and their probabilities
\cite{Po92}.

This work was supported by the Alexander von Humboldt
Foundation (YL), the Israel Academy of Sciences (YL and OE-W), the
German-Israeli Foundation and the Deutsche Forschungsgemeinschaft (PW).
We acknowledge helpful discussions with I. Ussishkin.

\end{multicols}
\end{document}